\documentstyle[epsfig,12pt,twoside]{article}

\newcommand{\be}{\begin{equation}}
\newcommand{\ee}{\end{equation}}
\newcommand{\bea}{\begin{eqnarray}}
\newcommand{\eea}{\end{eqnarray}}
\newcommand{\bean}{\begin{eqnarray*}}
\newcommand{\eean}{\end{eqnarray*}}

\begin{document}



\title{A New Decomposition of $\pi\pi$ S-wave Interaction
       \footnote{Supported in part by National Natural Science Foundation of 
                 China under Grant Nos.19775051 19875024 and 19905011}} 
\vspace{2cm}

\author{
Long Li${}^{2}$, Bing-Song Zou${}^{1,2}$, Guang-lie Li${}^{1,2}$
\footnote{Email: lilong@alpha02.ihep.ac.cn, zoubs@alpha02.ihep.ac.cn,
                 ligl@alpha02.ihep.ac.cn} \\
${}^1$ CCAST(World Laboratory), P.O.Box 8730, Beijing 100080, China\\
${}^2$ Institute of High Energy Physics, P.O.Box 918(4), Beijing 100039, China
 \footnote{Mailing address}}
\date{\today}

\maketitle

\begin{abstract}
  The experimental isoscalar $\pi\pi\to\pi\pi$ S-wave amplitude squared
below 1.75 GeV is characterized by a very broad structure
$f_0(400$-$1200)$ with two narrow dips due to its destructive interference
with the $f_0(980)$ and $f_0(1500)$. The $f_0(1370)$ and $f_0(1710)$
do not show up clearly in the $\pi\pi\to\pi\pi$ S-wave amplitude due to
their weak coupling to $\pi\pi$. This paper is about the controversial
nature of the broad $f_0(400$-$1200)$. We decompose it into two parts,
{\sl i.e.,} $t$-channel $\rho$ meson exchange plus an additional
$s$-channel resonance $f_0(X)$. With the $t$-channel $\rho$ meson exchange
fixed by the isotensor $\pi\pi\to\pi\pi$ S-wave scattering, we re-fit the
CERN-Munich (CM) data on $\pi\pi$ scattering to get parameters for the
$f_0(X)$. We find that the $f_0(X)$ is very broad with a nearby pole at
$(1.67-0.26i)$ GeV, while the $t$-channel $\rho$ meson exchange
part produces a pole at $(0.36-0.53i)$ GeV.
The implication of our results for the study of $\sigma$ and glueball is 
discussed.
\end{abstract}
\vspace{1cm}
{\bf PACS number(s): 14.40Cs, 11.80.Gw, 13.75.Lb}

\newpage

\section{Introduction}

As stated by the Particle Data Group\cite{PDG00}, the $I=0$
$J^{PC}=0^{++}$ sector is the most complex one, both experimentally and
theoretically, in the light quark meson spectroscopy. Meanwhile it is
also the most important and intriguing sector. It has the quantum number
of the vacuum, the $\sigma$ meson and the lightest glueball. The
importance of the $\sigma$ in relation to the dynamical chiral symmetry
breaking\cite{sigma1} and in reproducing the strong interaction between
nucleons\cite{sigma2} has long been known. 
The lightest glueball are predicted to be around $1.5 GeV$ 
by various lattice QCD calculations\cite{Lattice} and all sorts of QCD
inspired models\cite{QCDModel}. 
However the identification of both $\sigma$ and glueball is still not
settled.

A good place to study the properties of these isoscalar $0^{++}$ particles
is the $\pi\pi$ S-wave scattering process, since if a $0^{++}$ resonance
has a substantial coupling to $\pi\pi$, then it should show up clearly in
the $\pi\pi\to\pi\pi$ S-wave amplitude. In this context, the
$\pi^+\pi^-\to\pi^+\pi^-$ scattering from the old $\pi N$ scattering
experiments with both unpolarized\cite{CM} and polarized
targets\cite{Becker} has been re-analyzed\cite{BSZ, Kaminski} in
combination with new information from $p\bar p$ and other experiments.
The $\pi^+\pi^-\to\pi^0\pi^0$ scattering has also been studied by
E852\cite{E852} and GAMS\cite{GAMS} Collaborations recently.
All these efforts result in a consistent picture for the isoscalar
$\pi\pi$ S-wave intensity, i.e., a broad structure $f_0(400$-$1200)$ with
two narrow dips due to its destructive interference with the $f_0(980)$
and $f_0(1500)$.  The $f_0(1370)$ and $f_0(1710)$
do not show up clearly in the $\pi\pi\to\pi\pi$ S-wave amplitude due to 
their weak coupling to $\pi\pi$\cite{BSZ,WA102}.

Although the existence of the broad $f_0(400$-$1200)$ resonance is
experimentally well established\cite{PDG00}, its nature is still far
from settled. Some people believe that it is an intrinsic pole due to  
$q\bar q$ $\sigma$ meson\cite{Tornqvist,Ishida} or
glueball\cite{Anisovich,Ochs} while others claim that it is a dynamical
pole mainly due to t-channel exchange
forces\cite{Weinstein,Speth,Zou94,Locher,Oset}. A criticism\cite{Isgur} on
the intrinsic pole approaches is their neglect of the``exotic" isotensor
$\pi\pi$ scattering channel. In the dynamical pole approaches, the
isotensor $\pi\pi$ scattering channel can be used to set the scale of
t-channel forces. 

However, in the t-channel exchange approaches\cite{Weinstein,Speth} with
constraints from the isotensor $\pi\pi$ scattering channel, the t-channel
exchanges alone under-estimate the broad structure in the isoscalar
$\pi\pi$ scattering channel. Some addition S-channel resonance 
contribution is needed to full reproduce the broad structure.
Refs.\cite{Weinstein,Speth} attribute it to the tails of higher resonances
such as $f_0(1370)$ and $f_0(1500)$, while others\cite{Tornqvist,Igi}
believe that an intrinsic $q\bar q$ $\sigma$ is necessary.     

From information of other sources\cite{WA102,CB4pi}, we know that
the $f_0(1370)$ couples very weakly to $\pi\pi$ and the $f_0(1500)$ is
rather narrow. Both are unlikely to compensate the discrepancy between
the t-channel exchange calculations\cite{Weinstein,Speth} and the data.
The t-channel exchange calculation of Ref.\cite{Zou94} without any free
parameter and form factors reproduce the isoscalar $\pi\pi$ S-wave
phase shifts very well, but over-estimate the isotensor $\pi\pi$ S-wave
phase shifts\cite{Locher}. The off-shell form factors for the t-channel
exchange mesons is necessary to reproduce the isotensor $\pi\pi$ S-wave
data\cite{Locher} and will result in an under-estimation of isoscalar
$\pi\pi$ S-wave phase shifts as in Refs.\cite{Weinstein,Speth}.
Some additional s-channel resonance seems necessary.

In this paper, we decompose the broad $f_0(400$-$1200)$ structure into two
parts, {\sl i.e.,} $t$-channel $\rho$ meson exchange plus an additional
$s$-channel resonance $f_0(X)$. With the $t$-channel $\rho$ meson exchange
form factor parameter fixed by the isotensor $\pi\pi\to\pi\pi$ S-wave
scattering, we re-fit the old CERN-Munich (CM) data\cite{CM} on $\pi\pi$
scattering to get parameters for the $f_0(X)$. 

\section{Formalism}

\subsection{$t$-channel $\rho$ meson exchange amplitude}

We follow the K-matrix formalism as in Refs.\cite{Zou94,Locher}. 
In order to explain the $\pi\pi\ I=2\ S$-wave experimental data, a form factor
is needed to take into account the off-shell behavior of the
exchanged mesons. We use a form factor of conventional monopole type at
each vertex:
\be
F(q^2)=\frac{\Lambda^2-m^2}{\Lambda^2+q^2},
\ee
where m and q is the mass and four-vector momentum of exchanged mesons, 
and $\Lambda$ is the cutoff parameter which will be determined by
experimental data. Then the Born term for the $\rho$ exchange is:

\bea
T^{Born}(I=0)&=&2G[(\frac{\Lambda^2-m_\rho^2}{\Lambda^2-t})^2
           \frac{s-u}{m_\rho^2-t}+(\frac{\Lambda^2-m_\rho^2}{\Lambda^2-u})^2
           \frac{s-t}{m_\rho^2-u}],\\
T^{Born}(I=2)&=&-{1\over 2}T^{Born}(I=0) 
\eea
where $s,t,u$ are the usual Mandelstam variables and 
$G=g^2_{\rho\pi\pi}/32\pi$=0.364.
Their S-wave projections are
\bea
K^{I\!=\!0}_S(s)&\!=\!&4G\{(\frac{m^2_\rho}{\Lambda^2}-1) 
\frac{\Lambda^2+2s-4m^2_\pi}{\Lambda^2+s-4m^2_\pi}
+\frac{2s+m^2_\rho-4m^2_{\pi}}{s-4m^2_{\pi}}ln{\frac{(s+m^2_\rho-4m^2_\pi)
\Lambda^2}{(s+\Lambda^2-4m^2_\pi)m^2_\rho}}\},\nonumber\\
& & \\
K^{I\!=\!2}_S(s)&\!=\!&-{1\over 2}K^{I=0}_S(s).
\eea
K-matrix unitarization is introduced by
\be
T^I_S(s)=\frac{K^I_S(s)}{1-i\rho_1(s)K^I_S(s)},
\ee
where $\rho_1(s)=(1-4m^2_{\pi}/s)^{1/2}$ is the phase space factor. The
relation between the S-wave amplitude and the phase shift parameters
$\delta_I$ and $\eta_I$ is
\be
T^I(s)=\frac{\eta_I(s)e^{2i\delta_I(s)}-1}{2i\rho_1(s)} .
\ee

\begin{figure}[htbp]
\vspace{-1.6cm}
\begin{center}
\includegraphics[width=12cm,height=16cm]{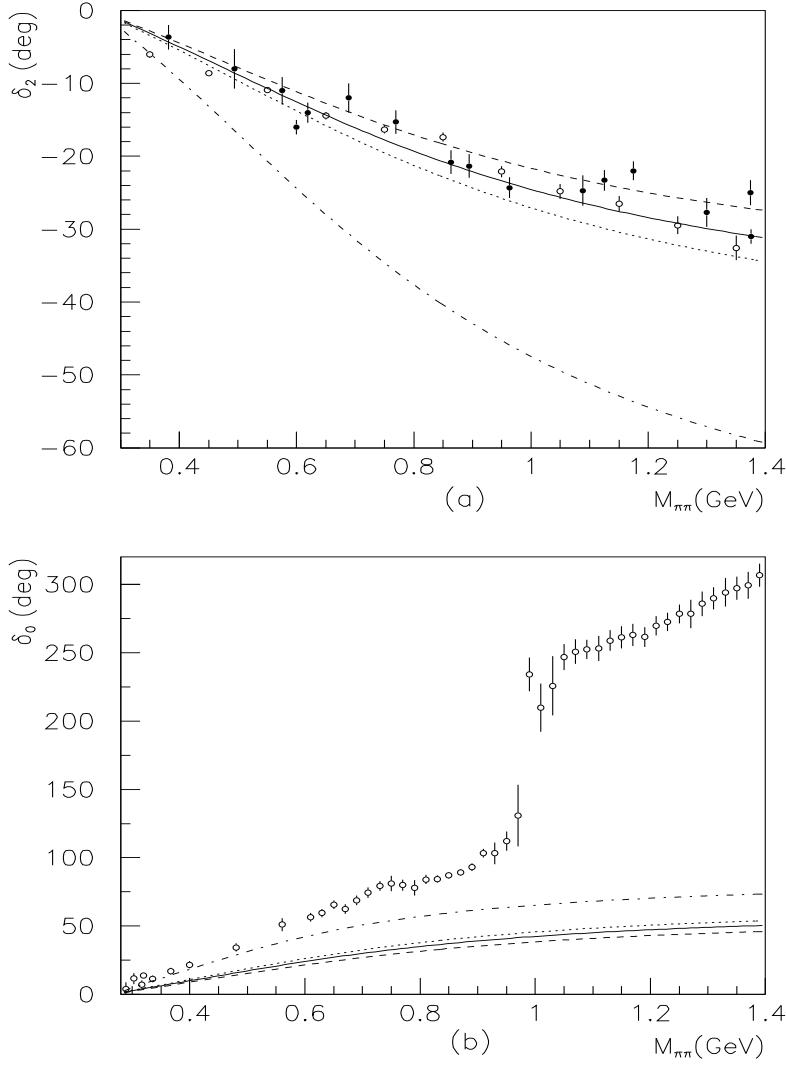}
\caption{The $I=2, 0$ $S$-wave phase shift $\delta_2$ (a) and $\delta_0$
(b) for $\pi\pi$ scattering. The experimental data for $\delta_2$ are
from Ref.\cite{Hoogland77} (circles) and Ref.\cite{Martin77}(dots);
The data for $\delta_0$ are from Ref.\cite{CM,Ke4}. 
The theoretical curves are from our t-channel $\rho$ exchange calculations
with $\Lambda=1400$ MeV (dotted lines), $\Lambda=1500$ MeV (solid lines),
$\Lambda=1600$ MeV (dashed lines) and without form factors (dot-dashed
lines). }
\label{fig:pp}
\end{center}
\end{figure}

From above formalism, we get the $\pi\pi\ I=2\ S$-wave phase shift 
$\delta_2(s)$ as shown in Fig.\ref{fig:pp}(a) for a few choices of
$\Lambda$ parameter. The dot-dashed curve without introducing the form
factor obviously overestimate the isotensor $\pi\pi$ S-wave phase shifts.
As in Ref.\cite{Locher}, the solid line with $\Lambda=1500$ MeV
reproduces the experimental isotensor $\pi\pi$ S-wave phase shifts best.
We also investigated the contribution from t-channel $f_2(1275)$ exchange.
It gives marginal improvement for the fit to the data with a form
factor cutoff parameter around 1 GeV as in Ref.\cite{Speth}.
Hence we do not include this contribution in our further calculations. 

Fig.\ref{fig:pp}(b) shows the corresponding contribution of $\rho$
exchange to the $I=0$ $S$-wave phase shift. Including the form factor
significantly reduces the t-channel $\rho$ exchange contribution.
The $t$-channel $\rho$ meson exchange
amplitude has a pole at $(0.36-0.53i)$ GeV.

With the $t$-channel $\rho$ 
meson exchange fixed by the isotensor $\pi\pi\to\pi\pi$ S-wave scattering, we 
will re-fit the CERN-Munich data to get information for s-channel
resonances.

\subsection{Full partial-wave amplitudes}

For each partial wave, it usually contains more than one components.
Specifically for isoscalar $\pi\pi$ S-wave, besides the t-channel $\rho$
exchange contribution, there are also contributions from several s-channel
resonances. To combine several components in a single partial wave, we use
the Dalitz-Tuan prescription\cite{BSZ}.
Suppose two components $a$
and $b$ for the partial wave $l$ are expressed individually as
$$T^a_l(s)=\frac{G_{a}}{C_a(s)-iG_{a}\rho_1(s)} \textrm{   and   } 
T^b_l(s)=\frac{G_{b}}{C_b(s)-iG_{b}\rho_1(s)}, \nonumber$$ 
then the combined amplitude will be 
\bea
\hat{T}^{ab}_l(s)&=&\frac{G_{a}C_b(s)+G_{b}C_a(s)}{[C_a(s)-iG_{a}\rho_1(s)]
[C_b(s)-iG_{b}\rho_1(s)]} \\
&=&\frac{G_{ab}}{C_{ab}(s)-iG_{ab}\rho_1(s)},
\eea
with $C_{ab}(s)=C_a(s)C_b(s)-G_{a}G_{b}\rho^2_1(s)$, and 
$G_{ab}=G_{a}C_b(s)+G_{b}C_a(s)$. At either component a simple pole 
term survives. The amplitude is explicitly unitary, and the
denominator contains poles from both components. Further poles are added
one by one using the equations above.

In fitting CERN-M\"unich data, we use the same formulae as the approach A
in Ref.\cite{BSZ} except for $\pi\pi$ S-waves. For the isotensor $\pi\pi$
S-wave, Ref.\cite{BSZ} used a scattering length formula, here we use the
t-channel $\rho$ exchange amplitude; for the isoscalar $\pi\pi$
S-wave, Ref.\cite{BSZ} used an effective S-channel $\sigma$ resonance to
describe the $f_0(400$-$1200)$ broad structure, here we decompose it into
two components, {\sl i.e.}, t-channel $\rho$ exchange plus an S-channel
resonance $f_0(X)$.

\section{Numerical results and discussion}

\begin{figure}[htbp]
\vspace{-1.6cm}
\begin{center}
\includegraphics[scale=0.7]{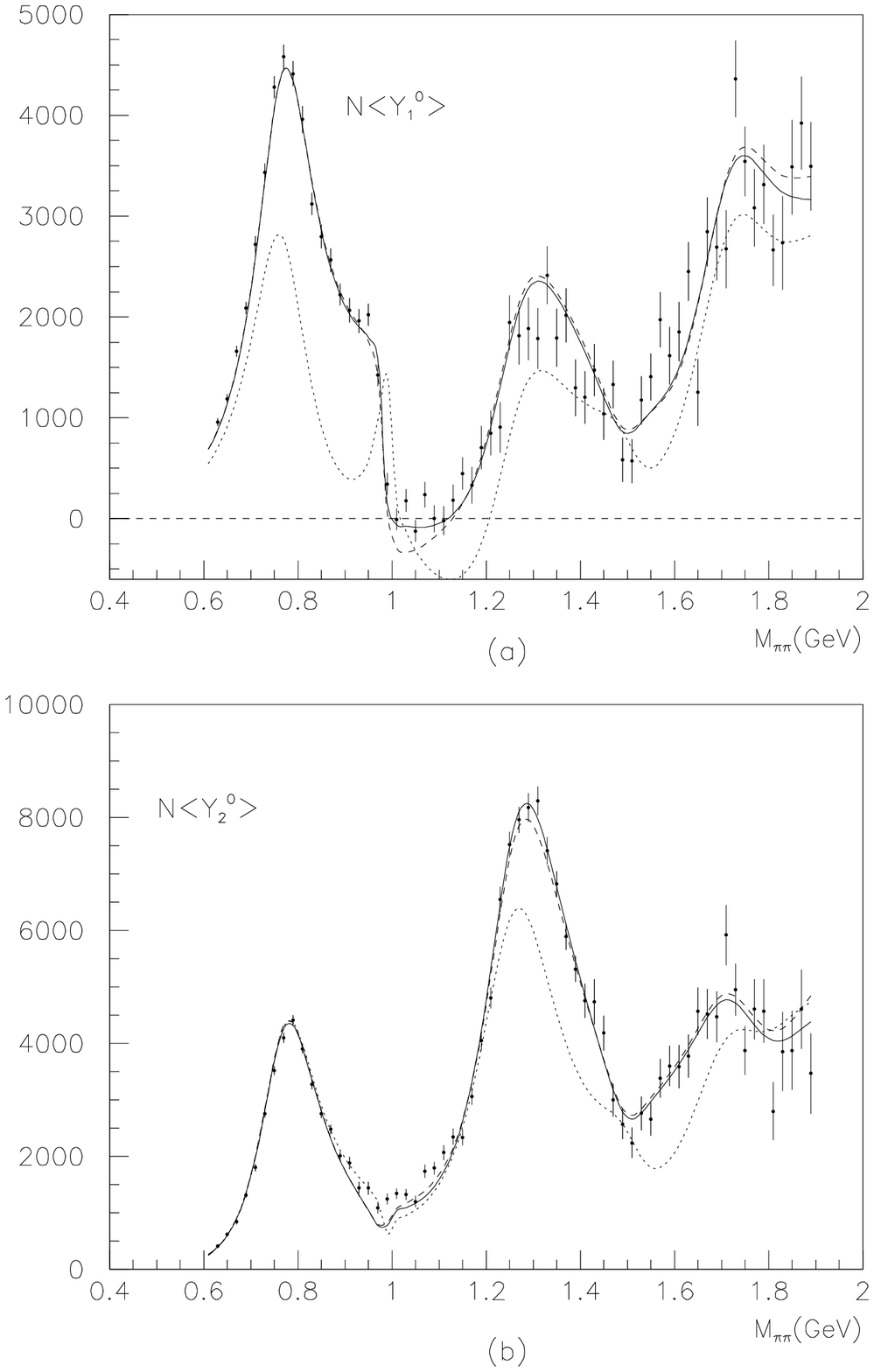}
\caption{The fits to CM data, 
the dashed line is the previous solution of Ref.\cite{BSZ}, 
the solid line is our fit and the dotted
line is for neglecting the contribution of $f_0(1670)$.} \label{cern}
\end{center}
\end{figure}

\begin{figure}[htbp]
\vspace{-0.6cm}
\begin{center}
\includegraphics[scale=0.7]{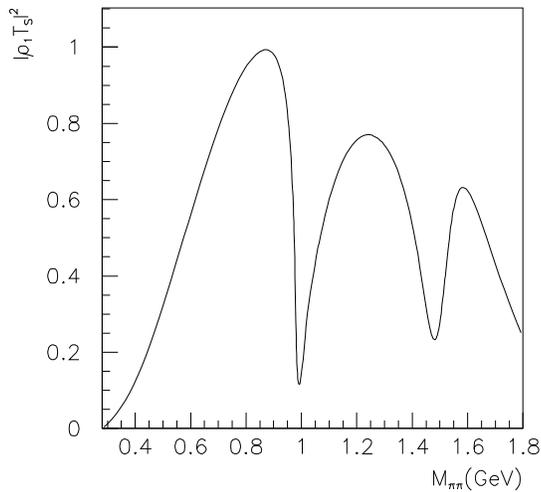}
\caption{The $I=0$ $\pi\pi$ S-wave amplitude squared.}
\label{sampl}
\end{center}
\end{figure}

With the t-channel $\rho$ exchange parameter $\Lambda=1.5 GeV$ fixed by
the isotensor $\pi\pi$ phase shifts, we re-fit the CM data on
the moments of the $\pi\pi$ decay angular distributions. The parameters
for relevant  resonances are fixed to the values of Ref.\cite{BSZ} except
for $f_0(X)$ and $f_0(980)$ which are responsible for the low energy
isoscalar $\pi\pi$ S-wave together with the t-channel $\rho$ exchange.
Among the moments of the CM data, the $N<Y_1^0>$ and $N<Y_2^0>$\cite{BSZ} 
are the most sensitive ones to the isoscalar $\pi\pi$ S-wave.   
Fig.~\ref{cern} shows the CM data, the previous solution of Ref.\cite{BSZ} 
(dashed lines) and our fit (solid lines) to these
moments.  The fitted parameters for the
$f_0(X)$ are $M_R=1.71$ GeV, $\Gamma_{\pi\pi}=1.12$ GeV and
$\Gamma_{4\pi}=0.16$ GeV which simulates all inelastic channels as in
Ref.\cite{BSZ} for the effective broad $\sigma$.
This $f_0(X)$ has a nearby pole at $(1.67-0.26i)$ GeV, so we denote it 
as $f_0(1670)$. 
As a comparison, we also show the results without the $f_0(1670)$
by the dashed lines in Fig.~\ref{cern}. It obviously fails to reproduce
data.
The fitted parameters for the
$f_0(980)$ are $M_R=0.982$ GeV, $g_{\pi\pi}=0.146$ GeV and
$g_{K\bar K}=0.562$ GeV in its Flatte formula
$1/(M^2_R-s-ig_{\pi\pi}\rho_{\pi\pi}-ig_{K\bar K}\rho_{K\bar K})$.
The corresponding full isoscalar $\pi\pi$ S-wave amplitude squared is
shown in Fig.\ref{sampl}. It is characterized by a very broad structure
with two narrow dips due to its destructive interference
with the $f_0(980)$ and $f_0(1500)$. The other two PDG
established $0^{++}$ resonances, $f_0(1370)$ and $f_0(1710)$,
do not show up clearly in the $\pi\pi\to\pi\pi$ S-wave amplitude due to 
their weak coupling to $\pi\pi$\cite{WA102}.

In summary, besides the $\sigma(400)$ pole at $(0.36-0.53i)$ GeV produced
by the t-channel $\rho$ exchange, there are five s-channel resonances
below 1.75 GeV:
four relative narrow PDG established ones, $f_0(980)$, $f_0(1370)$,
$f_0(1500)$ and $f_0(1710)$, plus a very broad one found here, $f_0(1670)$
with a nearby pole at $(1.67-0.26i)$ GeV. Theoretically there could be two 
resonances from $1^3P_0$
$q\bar q$ nonet, two from $2^3P_0$ nonet, then an extra from the
lightest glueball\cite{Anisovich}. The bare glueball state is dispersed
over three real resonances, $f_0(1500)$, $f_0(1710)$ and $f_0(1670)$.
This may give a nature explanation that $f_0(1500)$ and $f_0(1710)$ are
produced very strongly in the glue-rich process $J/\Psi$ radiative
decays\cite{PDG00,BES}. The $f_0(1670)$ may not be easy to be identified in
production processes due to its large decay width.

Note that glueballs could be very broad\cite{BPZ}. Lattice
QCD\cite{Lattice} predicts the $\pi\pi$ decay width alone of the lightest
glueball to be $(108\pm 29)$ MeV. Considering its possible larger decay
width to $\sigma\sigma$, $\rho\rho$ etc., the bare glueball state could
already have very large decay width. The mixing of overlapping resonances
could further increase the width of one state and simultaneously reduce
the width of another one\cite{Anisovich}. 

\vspace{1cm}

We thank D.V.Bugg and H.Q.Zheng for useful discussions.

\end{document}